\documentclass[aps,pra,superscriptaddress,twocolumn]{revtex4-1} % for PRL format
\usepackage{amsmath}
\usepackage{epstopdf}
\usepackage{graphicx}
\usepackage{bm,color,bbm}
\usepackage{caption}
\usepackage{subcaption}
\usepackage{hyperref}

\newcommand{\beq}{\begin{equation}}
\newcommand{\eeq}{\end{equation}}
\newcommand{\bqa}{\begin{eqnarray}}
\newcommand{\eqa}{\end{eqnarray}}
\newcommand{\nn}{\nonumber}

\begin{document}

\preprint{}

\title{Uncertainty relation for mutual information}


\author{James Schneeloch}
\affiliation{Department of Physics and Astronomy, University of Rochester, Rochester, New York 14627}
\affiliation{Center for Coherence and Quantum Optics, University of Rochester, Rochester, New York 14627}

\author{Curtis J. Broadbent}
\affiliation{Department of Physics and Astronomy, University of Rochester, Rochester, New York 14627}
\affiliation{Center for Coherence and Quantum Optics, University of Rochester, Rochester, New York 14627}
\affiliation{Rochester Theory Center, University of Rochester, Rochester, New York 14627}
\author{John C.  Howell}
\affiliation{Department of Physics and Astronomy, University of Rochester, Rochester, New York 14627}
\affiliation{Center for Coherence and Quantum Optics, University of Rochester, Rochester, New York 14627}

%\email[]{jschneel@pas.rochester.edu, curtis@pas.rochester.edu}
%\thanks{}
%\affiliation{University of Rochester, Dept. of Physics and Astronomy}


\date{\today}

\begin{abstract}
We postulate the existence of a universal uncertainty relation between the quantum and classical mutual informations between pairs of quantum systems. Specifically, we propose that the sum of the classical mutual information, determined by two mutually unbiased pairs of observables, never exceeds the quantum mutual information. We call this the complementary-quantum correlation (CQC) relation and prove its validity for pure states, for states with one maximally mixed subsystem, and for all states when one measurement is minimally disturbing. We provide results of a Monte Carlo simulation suggesting that the CQC relation is generally valid. Importantly, we also show that the CQC relation represents an improvement to an entropic uncertainty principle in the presence of a quantum memory, and that it can be used to verify an achievable secret key rate in the quantum one-time pad cryptographic protocol.
\end{abstract}

%\keywords{}

\maketitle


\section{Introduction}
Between a pair of quantum systems, the classical correlations (as quantified by the classical mutual information) never exceed the total intrinsic correlations (as quantified by the quantum mutual information)\cite{nielsen2000}. This idea is utilized extensively in studies of quantum discord
\cite{ollivierquantumdiscord,hendersonvedraldiscordpaper}, a measure of the correlations between a pair of quantum systems that cannot be accessed through any set of local measurements. Because quantum information leads to many useful applications including quantum computing \cite{Deutsch1985QC}, quantum cryptography \cite{bb84paper,Scarani2009}, and quantum superdense coding \cite{BennetSuperdense1992}, understanding quantum correlations is especially important. Finding tight bounds to the quantum mutual information with classical measurements is particularly useful, since direct calculation of the quantum mutual information requires complete knowledge of the density operator, which is impractically difficult to obtain for higher-dimensional systems. In addition, entanglement criteria and security bounds in quantum cryptography often rely on the strength of the fundamental quantum correlations given by the quantum mutual information \cite{Scarani2009,Schumacher_QMI1TP_PRA2006}; if one can provide a tight lower bound to the quantum mutual information, verifying security in quantum key distribution (QKD) and witnessing entanglement can become considerably easier.

In this article, we examine the relationship between the quantum mutual information of a joint $M\otimes N$ system $AB$ before and after a set of joint local projective measurements, say of observables $\hat{Q}^{A}$ and $\hat{Q}^{B}$ of systems $A$ and $B$, respectively. This post-measurement quantum mutual information is equal to the classical mutual information obtained from the joint probability distribution of measurement outcomes \cite{discordRMPModi2012}, $P(q^{A}_{i},q^{B}_{j})$, where indices $i$ and $j$ run over all measurement outcomes.

While it is known that the classical correlations between $\hat{Q}^{A}$ and $\hat{Q}^{B}$ never exceed the total quantum correlations in $AB$ \cite{nielsen2000}, we show that a tighter bound can be obtained in many cases, and we provide evidence which suggests that the tightened bound may be universally valid for $M\otimes N$ systems. In these cases, the sum of correlations between one pair of observables (say, $\hat{Q}^{A}$ and $\hat{Q}^{B}$) and the correlations between a second complementary (or mutually unbiased) pair of observables ($\hat{R}^{A}$ and $\hat{R}^{B}$) never exceeds the total quantum correlations \footnote{What we mean by the classical correlations between $\hat{Q}^{A}$ and $\hat{Q}^{B}$ is the classical mutual information of the joint probability distribution of the measurement outcomes of $\hat{Q}^{A}$ and $\hat{Q}^{B}$. What we mean by the total quantum correlations is simply the quantum mutual information of the joint system $AB$. This is not to be confused with measures of the quantumness of correlations (i.e., the quantum discord).} in the bipartite quantum system. We call this relationship \eqref{bigresult} between the complementary correlations and the quantum correlations the \emph{complementary-quantum correlation} (CQC) relation;
\begin{equation}\label{bigresult}
H(\hat{Q}^{A}\!:\!\hat{Q}^{B}) + H(\hat{R}^{A}\!:\!\hat{R}^{B})\leq I(A\!:\!B),
\end{equation}
where $I(A\!:\!B)$ is the quantum mutual information of density operator $\hat{\rho}^{AB}$ (the state of system $AB$ before measurement), $I(A\!:\!B)\equiv S(A)+S(B)-S(A,B)$; $H(\hat{Q}^{A}\!:\!\hat{Q}^{B})$ is the classical mutual information obtained from the joint distribution of measuring $\hat{Q}^{A}$ of $A$ and $\hat{Q}^{B}$ of $B$; and $H(\hat{R}^{A}\!:\!\hat{R}^{B})$ is similarly defined for observables $\hat{R}^{A}$ and $\hat{R}^{B}$, respectively mutually unbiased with $\hat{Q}^{A}$ and $\hat{Q}^{B}$ \footnote{By mutually unbiased, we mean that a system $A$ in an eigenstate of $\hat{Q}^{A}$ is equally likely to be measured in any of the eigenstates of $\hat{R}^{A}$ (i.e., that they are maximally uncertain with respect to one another), and vice versa.}. Note here that $\hat{Q}^{A}$ and $\hat{Q}^{B}$ are an arbitrary pair of observables, and that we only require that $\hat{R}^{A}$ be mutually unbiased with $\hat{Q}^{A}$, and that $\hat{R}^{B}$ be mutually unbiased with $\hat{Q}^{B}$. In addition, the CQC relation can be regarded as an uncertainty relation for mutual information since, like every other uncertainty relation, it is a constraint on the measurement probabilities of quantum systems.

In the following, we provide proofs of this CQC relation for all pure discrete bipartite systems, for all discrete bipartite systems in which one of the subsystems is maximally mixed, and for all bipartite systems when one of the pairs of observables minimally disturbs the system \footnote{A minimally disturbing measurement $\hat{Q}$ is one in which $\hat{Q}$ commutes with the state (or reduced state, in this case) under measurement.}. In addition, we prove the CQC relation by direct calculation for asymmetric Werner states \cite{Werner1989} of two-qubit systems. We next show evidence that the CQC-relation \eqref{bigresult} is satisfied in general using Monte Carlo simulations of random bipartite states of dimension up to ($4$ by $4$). We then show that the CQC relation can be viewed as an improvement on an entropic uncertainty principle in the presence of a quantum memory \cite{Berta2010} by demonstrating its improvement using asymmetric Werner states. Finally, we show that when coupled with the strong subadditivity of the entropy \cite{nielsen2000}, the CQC relation can be used to determine a lower bound to the secret key capacity in the quantum one-time pad cryptographic protocol \cite{Schumacher_QMI1TP_PRA2006}.

\section{Proof of the CQC relation for pure states}
Proving the CQC relation for pure discrete bipartite states can be done using the work of Hall \emph{et~al.} \cite{Hall2006} and Luo \emph{et~al.} \cite{Luo2009}, who showed that for pure states, the quantum mutual information is never less than twice the classical mutual information;
\begin{equation}
I(A\!:\!B) = 2\max_{\{\hat{Q}^{A},\hat{Q}^{B}\}}\;I(\hat{Q}^{A}\!:\!\hat{Q}^{B}).
\end{equation}
Knowing this, the quantum mutual information  for pure states is never less than the sum of any two classical mutual informations of any two pairs of observables (not just mutually unbiased pairs):
\begin{equation}
I(A:B)\geq H(\hat{Q}^{A}\!:\hat{Q}^{B}) + H(\hat{R}^{A}\!:\hat{R}^{B}),
\end{equation}
proving our CQC relation \eqref{bigresult} for all  pure discrete bipartite quantum systems. We note that it is readily seen that \emph{this} CQC relation cannot be applied to arbitrary observable pairs for mixed states because the quantum mutual information is not always larger than twice any classical mutual information \footnote{The notion that the quantum mutual information never exceeds twice the largest possible classical correlations (i.e., the Lindblad conjecture \cite{LindbladConj}) was disproved by Luo \emph{et~al.} \cite{Luo2009}.}. In addition, the mutual information is neither convex nor concave in the joint distribution or density matrix. This prevents us from concluding that the CQC relation is valid for mixed states from its validity for pure states. Finding similar uncertainty relations which accommodate arbitrary pairs of observables is the subject of ongoing investigation.

\section{Proof of the CQC relation for zero residual uncertainty}
Proving that the CQC relation is valid for bipartite systems with at least one maximally mixed subsystem, and when one of the measurements is minimally disturbing, follows as a special case of Berta \emph{et~al.}'s uncertainty principle in the presence of quantum memory \cite{Berta2010}. The following uncertainty relation (for mutually unbiased $\hat{Q}^{A}$ and $\hat{R}^{A}$),
\begin{equation}\label{BertaLoose}
H(\hat{Q}^{A}|\hat{Q}^{B}) + H(\hat{R}^{A}|\hat{R}^{B})\geq \log(N^{A}) + S(A|B),
\end{equation}
[where $S(A|B) \equiv S(A,B) - S(B)$, and $N^{A}$ is the dimensionality of $A$] is a classical version of Berta \emph{et~al.}'s uncertainty principle in the presence of quantum memory \cite{Berta2010}, already known to be true. Note that here and throughout the paper, all logarithms are base $2$ unless otherwise specified. We can then express our CQC relation in terms of the entropies seen in Berta's relation \eqref{BertaLoose}:
\begin{align}\label{CQCexpanded2}
H(\hat{Q}^{A}&|\hat{Q}^{B}) + H(\hat{R}^{A}|\hat{R}^{B})\geq \log(N^{A}) + S(A|B)\nn\\
 &+[ H(\hat{Q}^{A}) + H(\hat{R}^{A}) - \log(N^{A}) -S(A)].
\end{align}
Here, we see that the CQC relation represents an improvement to Berta's uncertainty relation, by raising the uncertainty limit by the amount of the last four terms in brackets. The sum in brackets (which we define as the residual uncertainty of system $A$) is always non-negative. It can be shown that the terms in brackets are an expression of Berta's uncertainty relation when systems $A$ and $B$ are completely uncorrelated from one another. 

To show that the CQC relation is valid when a subsystem is maximally mixed, or when a set of measurements is minimally disturbing, we note that whenever the residual uncertainty of $A$ is zero, our CQC relation is equivalent to Berta's uncertainty relation, and therefore valid. When system $A$ is maximally mixed, $H(\hat{Q}^{A})$, $H(\hat{R}^{A})$, and $S(A)$ are all equal to $\log(N^{A})$, making the residual uncertainty zero. When $\hat{Q}^{A}$ [alternatively, $\hat{R}^{A}$] minimally disturbs system $A$, $H(\hat{Q}^{A})$ [alternatively, $H(\hat{R}^{A})$] is equal to $S(A)$, and  $H(\hat{R}^{A})$ [alternatively, $H(\hat{Q}^{A})$] is equal to $\log(N^{A})$, which again, makes the residual uncertainty zero. Since we may switch parties $A$ and $B$ when writing Eqs.~\eqref{BertaLoose} and \eqref{CQCexpanded2}, the CQC relation (already symmetric between parties) is valid when either $A$ or $B$ is maximally mixed, and when either $\hat{Q}^{A}$, $\hat{Q}^{B}$, $\hat{R}^{A}$, or $\hat{R}^{B}$ is a minimally disturbing measurement. These results extend the validity of the CQC relation to  a wide variety of states, including all Bell-diagonal states, Werner states, and any maximally correlated mixed states.

\begin{figure}[t]
 \centering
\includegraphics[width=\columnwidth]{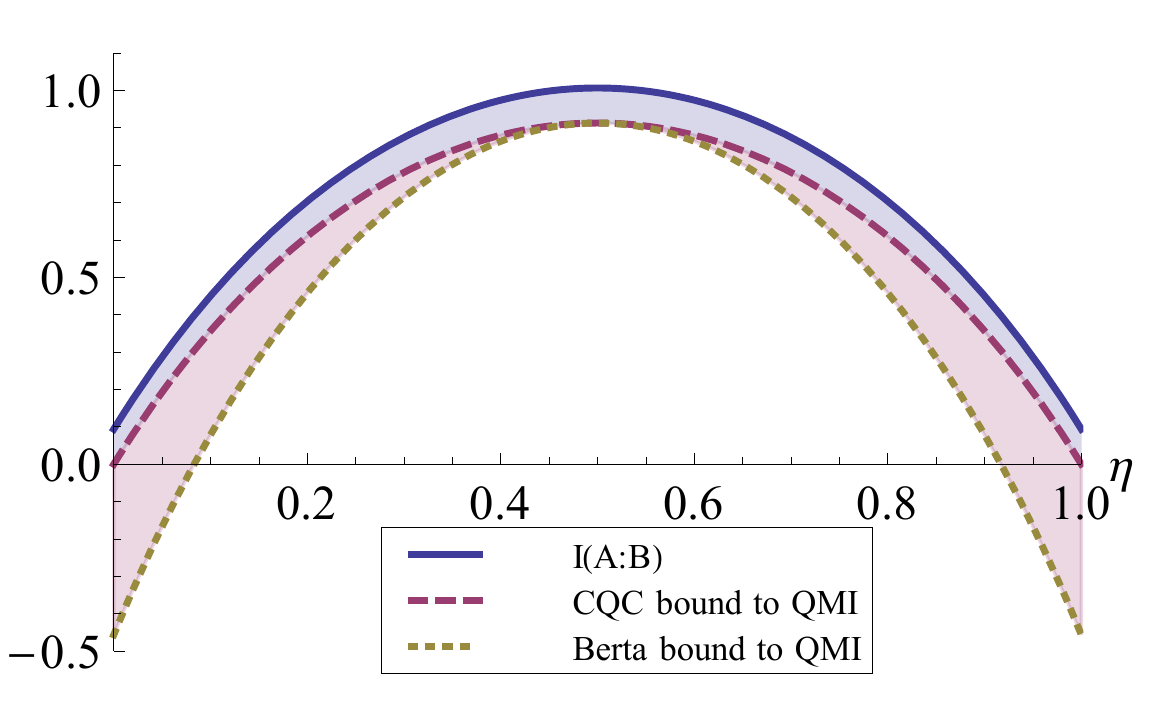}
\caption{(Color online) The solid curve is a plot of the quantum mutual information $I(A:B)$ for the asymmetric Werner state \eqref{asymwerner} with $p=3/4$. The dashed curve is the CQC bound [i.e., $H(\hat{\sigma}_{X}^{A}:\hat{\sigma}_{X}^{B})$ + $H(\hat{\sigma}_{Y}^{A}:\hat{\sigma}_{Y}^{B})$] to the quantum mutual information for this state, and the dotted curve is the bound obtained from the classical version of Berta's uncertainty relation [i.e., $S(A)$ + $\log(N^{A})$ - $H(\hat{\sigma}_{X}^{A}|\hat{\sigma}_{X}^{B})$ - $H(\hat{\sigma}_{Y}^{A}|\hat{\sigma}_{Y}^{B})$]. The shading highlights the difference between the various quantities. For $\eta$ near zero or unity, the CQC relation offers a substantial improvement over Berta's uncertainty relation.}
\end{figure}

\begin{figure*}[t]
 \centering
\includegraphics[width=\textwidth]{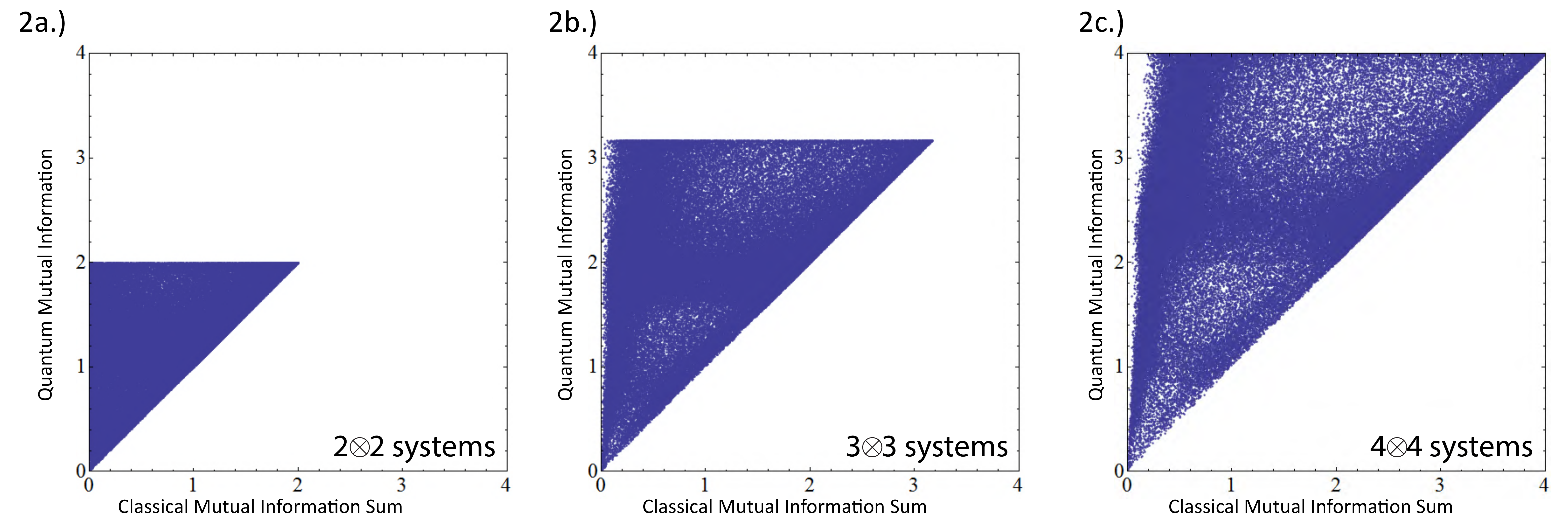}
\caption{(Color online) Scatterplots of the quantum mutual information as a function of the sum of the classical mutual informations determined by a pair of unbiased measurement bases. (a)-(c): Plots for our randomly perturbed $2\otimes 2$, $3\otimes 3$, and $4\otimes 4$ boundary states, respectively. The sharp diagonal boundary illustrates that perturbing boundary states does not yield states that violate our CQC relation.}
\end{figure*}

The CQC relation is also closely connected with information exclusion relations, first described by Hall \cite{Hall1995}. Indeed, when one of the subsystems is maximally mixed and the observables $\{\hat{Q}^{A},\hat{Q}^{B}\}$ and $\{\hat{R}^{A},\hat{R}^{B}\}$ are mutually unbiased, the information exclusion relation proven in \cite{ColesPianiInfExcRel2014}, as well as the uncertainty relation in \cite{Christandl2005} both reduce to the CQC relation \footnote{Marco Piani (2014, private communication)}.

In order to test the CQC relation for classes of states in which the residual uncertainty is not zero, we first considered the class of $2\otimes 2$ asymmetric Werner states denoted by $\hat{\rho}_{ASW}$;
\begin{align}\label{asymwerner}
\hat{\rho}_{ASW} &\equiv p|\psi^{-}_{AS}\rangle\langle\psi^{-}_{AS}|+(1-p) \frac{\mathbf{I}}{4},\nn\\
:&|\psi^{-}_{AS}\rangle\equiv \sqrt{\eta}\;|\uparrow,\downarrow\rangle -\sqrt{1-\eta}\;|\downarrow,\uparrow\rangle,
\end{align}
where $p$ and $\eta$ are real numbers between zero and unity. Here, $|\uparrow\rangle$ and $|\downarrow\rangle$ are, respectively, the $+1$ and $-1$ eigenstates of the Pauli $\hat{\sigma}_{Z}$ observable. As a function of $p$ and $\eta$, we calculated the classical mutual informations $H(\hat{\sigma}_{X}^{A}\!:\!\hat{\sigma}_{X}^{B})$ and $H(\hat{\sigma}_{Y}^{A}\!:\!\hat{\sigma}_{Y}^{B})$, and compared their sum to the quantum mutual information. As illustrated in Fig.~1, the CQC relation is everywhere satisfied for this class of states, and significantly tightens the bound obtained from the classical version of Berta's uncertainty relation \eqref{BertaLoose}.

\section{Numerical Investigation of the CQC relation for arbitrary states}
To investigate the general validity of the CQC relation, we simulated many random bipartite states in search of counterexamples. We tested the CQC relation on a uniform sampling of $10^{7}$ random: $2\otimes 2$ systems, $2\otimes 3$ systems, and $3\otimes 3$ systems, as well as $10^{6}$ random $2\otimes 4$ systems, $3\otimes 4$ systems, and $4\otimes 4$ systems using the algorithm in \cite{RandomDensityPRL1998}. For each bipartite state, we calculated the quantum mutual information, and the classical mutual information in two fixed pairs \footnote{Since the states generated randomly were rotated by a random unitary transformation, there is no need to also select a random set of mutually unbiased measurement bases} of mutually unbiased observables. These simulations yielded no counterexamples, and almost all randomly generated states had nonzero residual uncertainties, improving on Berta's uncertainty relation \eqref{BertaLoose}. In addition, using randomly generated unitary transformations, we perturbed $N\otimes N$ states known to saturate the CQC relation. These boundary states include mixtures of the symmetric Bell state, $|\Phi^{+}\rangle =\frac{1}{\sqrt{N}}\sum_{i=1}^{N}|i,i\rangle$, and the maximally correlated mixed state, $\hat{\rho}_{MCM} =\frac{1}{N}\sum_{i=1}^{N}|i,i\rangle\langle i,i|$, as well as mixtures of the maximally correlated mixed state with the maximally mixed state, $\hat{\rho}_{MM}=\frac{1}{N^{2}}\sum_{i,j=1}^{N}|i,j\rangle\langle i,j|$. The resulting states showed no violation. This provides strong supporting evidence for the general validity of the CQC relation \footnote{Additional numerical investigations would be required to arrive at a numerical proof.}. The plots in Fig.~2 show the results of the perturbed boundary state simulations, where we see both the sharp boundary that is the CQC-relation, and the upper limit of the quantum mutual information.

\section{Applications of the CQC relation}
If the CQC relation can be shown to be generally valid, it will have applications sprouting from two improved abilities: finding a lower limit to the quantum mutual information with the complementary classical mutual information sum, and in providing an upper limit to the complementary classical mutual information sum with the quantum mutual information. From the first ability, the CQC relation allows us to witness nonclassical values of the quantum mutual information, which is impossible when comparing only one classical mutual information to the quantum mutual information. Indeed, if the classical mutual information sum is larger than the least of the marginal classical entropies, then the joint system must be entangled [i.e., the conditional quantum entropy must be negative]. This is an improvement over using Berta's uncertainty relation, where the mutual information sum would need to be larger than the difference between the sum [$H(\hat{Q}^{A})+H(\hat{R}^{A})$] and $\log(N^{A})$. 

The latter ability allows us to place theoretical limits on our ability to witness entanglement and Einstein-Podolsky-Rosen (EPR) steering \cite{Wiseman2007,Cavalcanti2009}. If the quantum mutual information is less than the largest possible classical mutual information [i.e., $\log(N)$ in an $N\otimes N$ system], then we will not be able to demonstrate EPR-steering via the inequality in \cite{Schneeloch2013}, even with an optimal choice of complementary observables. This improves upon the prior result \cite{Berta2010}, in which a quantum mutual information less than either $S(A)$ or $S(B)$ is the limit which prevents us from violating the symmetric steering inequality in \cite{Schneeloch2013}. This does not, however, mean that $I(A\!:\!B)$ must exceed $\log(N)$ in an $N\otimes N$ system to exhibit symmetric steering. As has been proven \cite{Gisin1991}, any pure entangled state is Bell nonlocal, and so also symmetrically steerable, even for a quantum mutual information near zero.

To explore other applications of the CQC relation, we must consider the physical significance of the quantum mutual information. Among other applications, the quantum mutual information represents the channel capacity of the quantum one-time pad \cite{Schumacher_QMI1TP_PRA2006}, which is a secret quantum communication protocol (independent of QKD) similar to superdense coding. In the ideal implementation of the quantum one-time-pad \cite{Schumacher_QMI1TP_PRA2006}, Alice and Bob share a pair of qubits whose joint state is the Bell-singlet state \footnote{They do not need to share pairs in only the Bell-singlet state. The joint states only need to be prepared identically, so that Alice and Bob both know what the initial joint states are.}. To send a message, Alice performs on her qubit one of an agreed-upon set (i.e., alphabet) of possible unitary transformations. Alice's choice of transformation changes the joint state that she and Bob share to be any one of the four Bell states Alice chooses. When she sends her qubit to Bob, he can perform measurements on both qubits to identify the Bell state, and determine which transformation Alice performed. To send a long message, Alice and Bob would use many pairs of qubits since each pair's entanglement would be consumed in each use of the protocol.

The quantum one-time pad is more than simply quantum superdense coding, as the manipulations Alice performs must be restricted so that, on average, the quantum states she sends to Bob are indistinguishable from one another. In fact, the capacity for quantum superdense coding \cite{Bowen2001} exceeds the capacity for the quantum one-time pad, since one can optimize information transfer over all channel inputs without any restrictions.

The quantum one-time pad is particularly useful for two reasons. First, it offers an increased channel capacity similar to superdense coding, making it more efficient than a classical one-time pad. Second, no entangled pairs are discarded in a key-sifting step (as in QKD), making it more economical in terms of the physical resources required.

What makes the quantum one-time pad secure is that the reduced density matrices of all four Bell states are completely indistinguishable from one another. Thus, a third party adversary with unlimited classical resources at their disposal can do no better than random chance at guessing the message (just as in a classical one-time pad). This notion of security is distinct from the security in QKD protocols, which rely on the ability to detect eavesdropping. In the quantum one-time pad with a classical adversary, the worst an eavesdropper can do is to prevent information from being transmitted (i.e., a denial-of-service attack). The quantum one-time pad's security is similar to the security in the classical one-time pad; without knowledge of the key string, all possible ciphertexts are equally likely to be generated from the plaintext message because all possible key strings are equally likely (being randomly generated). In the quantum one-time pad, the quantum key string can be thought of as the qubits that Bob originally has that are entangled with Alice's qubits. In this way, the quantum and classical one-time pads share the same strengths and vulnerabilities. As long as the ``key" is shared exclusively between Alice and Bob, both protocols are secure against all cryptographic attacks.

In more realistic scenarios, the quantum mutual information between Alice's and Bob's subsystems was shown \cite{Schumacher_QMI1TP_PRA2006} to be an achievable asymptotic secret key rate between Alice and Bob when a third party, Eve, has unlimited classical resources at her disposal to intercept the subsystems and deduce Alice's manipulations, but has no additional side information. With this application in mind, the CQC relation allows two parties, Alice and Bob, to experimentally determine a lower limit to their quantum mutual information, and thus to their maximum key rate in the quantum one-time pad protocol with classically equipped adversaries.

In more general scenarios, where a classical adversary might actually have some information about the key string, it was shown in \cite{csiszar1978broadcast}, that an achievable secure key rate between Alice and Bob is equal to the difference between the mutual information between Alice's message string and Bob's decryption of it, and the mutual information between Alice's message string and Eve's best estimation of it \footnote{In \cite{devetak2005distillation}, this idea of side information was partially extended to the quantum regime, where instead Bob and Eve receive quantum systems that are classically correlated to Alice's key string, and must then determine what Alice's key string is through an optimal set of measurements. In this event, an achievable secret key rate between Alice and Bob was shown to be the difference between the quantum mutual information between Alice's key string and Bob's systems, and the quantum mutual information between Alice's key string and Eve's quantum systems.}. As we shall show, the CQC relation allows us to place a lower bound on a secret key rate to the quantum one-time pad, even when the adversary Eve has quantum side information (here meant as quantum systems partially entangled with Alice's qubits).

To find such a lower bound to the secret key rate in the quantum one-time pad, we use that the CQC relation allows Alice and Bob to experimentally determine (using a small fraction of their total pairs) an upper limit to the quantum mutual information that a third party, Eve, can share with either of them. From a particular formulation of the strong subadditivity of the entropy \cite{nielsen2000}, one can arrive at a weak monogamy inequality of nonclassical correlations. For any tripartite system shared by parties Alice, Bob, and Eve, described by density matrix $\hat{\rho}^{ABE}$,
\begin{equation}
I(A\!:\!B) + I(A\!:\!E)\leq 2 S(A)\leq 2\log(N^{A}).
\end{equation}
Combining this with our CQC-relation \eqref{bigresult}, we find that
\begin{equation}
I(A\!:\!E)\leq 2\log(N^{A}) - H(\hat{Q}^{A}\!:\!\hat{Q}^{B})-H(\hat{R}^{A}\!:\!\hat{R}^{B}).
\end{equation}
In other words, Alice and Bob's measured mutual informations give them a firm upper bound to Eve's quantum mutual information with Alice. If Alice and Bob's mutual information sum is larger than $\log(N^{A})$ [and $\log(N^{B})$], then Alice and Bob have not only demonstrated symmetric EPR steering \cite{Schneeloch2013}; they have verified that Eve's mutual information with either Alice or Bob must be less than $\log(N^{A})$ [and $\log(N^{B})$] as well.

By placing an upper limit on Eve's mutual information with both Alice and Bob, Alice and Bob can place a lower bound on $R$, i.e., the minimum difference between the quantum mutual information between Alice and Bob, and the quantum mutual information that Eve has between either Alice or Bob:
\begin{equation}
R\equiv \min_{X=\{A,B\}} \big(I(A:B)-I(X:E)\big).
\end{equation}
If $H(\hat{Q}^{A}\!:\!\hat{Q}^{B}) + H(\hat{R}^{A}\!:\!\hat{R}^{B})\geq \log(N) + \epsilon$ (where Alice and Bob share $N\otimes N$ systems), then $R\geq 2 \epsilon$. $R$ is a conservative asymptotic secret key rate for two-way communication between Alice and Bob using the quantum one-time pad protocol in which Eve may also possess quantum information about Alice's and Bob's systems. This can be shown by using Branda\~o \emph{et~al.}'s single letter formula for the quantum one-time pad with an eavesdropper\cite{Brandao_Q1TPEve_PRL2012,Brandao_QCandSupAct_IEEE2013}, where the secret key rate is shown to be equal to the difference between the mutual informations relating Alice's preparations to the joint states that Bob and Eve receive through a symmetric side channel. If the states Alice sends through the channel are classically indistinguishable, as in Schumacher's original quantum one-time pad paper \cite{Schumacher_QMI1TP_PRA2006}, then each mutual information reduces in the asymptotic limit to the quantum mutual information between Alice and Bob, and between Alice and Eve, respectively.

\section{Conclusion}
Quantum entropies are harder to measure than classical entropies since they require knowledge of the eigenvalue spectrum of the density operator, which can only be found by experimentally determining the entire density operator. The quantum mutual information is particularly difficult to determine for large-dimensional systems because of how the number of measurements for tomography scales as $N^{4}$ for an $N\otimes N$ joint quantum system. We have shown promising evidence for, and proven many cases of, a new lower bound to the quantum mutual information using significantly fewer classical measurements than it would take to determine the density operators. We have also shown that if the CQC relation is generally valid, then two parties can bound the mutual information that a third party might have with either of them, which will be of practical use in quantum cryptography applications.

We gratefully acknowledge insightful discussions with Gregory A. Howland, Marco Piani, and Patrick Coles, as well as support from DARPA-DSO InPho Grant No. W911NF-10-1-0404, and DARPA-DSO Grant No. W31P4Q-12-1-0015. CJB acknowledges additional support from ARO W911NF-09-1-0385 and NSF PHY-1203931.

\bibliography{EPRbib13}

\end{document}